\documentclass[twocolumn]{aastex631}
\usepackage{multirow}

\usepackage{hhline} 

\usepackage{amsmath}
\usepackage{bigints}

\begin{document}

\title{Fundamental-harmonic pairs of interplanetary type III radio bursts}

\correspondingauthor{Immanuel Christopher Jebaraj}
\email{immanuel.c.jebaraj@gmail.com}

\author[0000-0002-0606-7172]{Immanuel Christopher Jebaraj}
\affiliation{Space Research Laboratory,
            University of Turku, Turku, Finland\\}

\author[0000-0002-6809-6219]{Vladimir Krasnoselskikh}
\affiliation{LPC2E/CNRS, UMR 7328, 3A Avenue de la Recherche Scientifique, Orleans, France}
\affil{Space Sciences Laboratory, University of California, Berkeley, CA 94720-7450, USA}

\author[0000-0002-1573-7457]{Marc Pulupa}
\affil{Space Sciences Laboratory, University of California, Berkeley, CA 94720-7450, USA}

\author[0000-0003-1169-3722]{Jasmina Magdalenic}
\affiliation{Center for mathematical Plasma Astrophysics, Department of Mathematics, KU Leuven, Celestijnenlaan 200B, B-3001 Leuven, Belgium}
\affiliation{Solar-Terrestrial Centre of Excellence -- SIDC, Royal Observatory of Belgium, Avenue Circulaire 3, 1180 Uccle, Belgium}

\author[0000-0002-1989-3596]{Stuart D. Bale}
\affil{Physics Department, University of California, Berkeley, CA 94720-7300, USA}
\affil{Space Sciences Laboratory, University of California, Berkeley, CA 94720-7450, USA}

\begin{abstract}

Type III radio bursts are not only the most intense but also the most frequently observed solar radio bursts. However, a number of their defining features remain poorly understood. Observational limitations, such as a lack of sufficient spectral and temporal resolution, have hindered a full comprehension of the emission process, especially in the hecto-kilometric wavelengths. Of particular difficulty is the ability to detect the harmonics of type III radio bursts. Hereafter we report the first detailed observations of type III fundamental-harmonic pairs in the hecto-kilometric wavelengths, observed by the Parker Solar Probe. We present the statistical analysis of spectral characteristics and the polarization measurements of the fundamental-harmonic pairs. Additionally, we quantify various characteristic of the fundamental-harmonic pairs, such as the time-delay and time-profile asymmetry. Our report and preliminary analysis conclude that fundamental-harmonic pairs constitute a majority of all type III radio bursts observed during close encounters when the probe is in close proximity to the source region and propagation effects are less pronounced.



\end{abstract}

\keywords{}

\section{Introduction} \label{sec:intro}

Type III radio bursts are the most intense and well-observed radio emissions of solar origin in the solar system. They are the radio signatures of energetic electron beams accelerated at the Sun, that stream away into the heliosphere along open magnetic field lines \citep[][]{McLean85book}. In the dynamic radio spectrum, which presents intensity as a function of time and frequency, type III bursts are distinguished as rapidly drifting emissions. Although various aspects of type III bursts-ranging from electron beam evolution, their interaction with the background plasma, and subsequent electromagnetic emission—are still not entirely understood, it is generally accepted that they are generated through the plasma emission mechanism which is a two-step process \citep[][and references therein]{Ginzburg58,Melrose80}. Initially, streaming electrons interact with the background plasma, generating Langmuir waves close to the electron plasma frequency ($f_p$). In the second, non-linear stage, these Langmuir waves are partially converted to electromagnetic (EM) waves by wave-wave or wave-particle interactions. The resultant EM wave is emitted at either close to $f_p$ or its harmonics \citep[$nf_p$; $n^{th}$ harmonic of the plasma frequency, where $n=2,3,\ldots$,][]{Robinson98a, Robinson98b, Robinson98c}. 

Type III radio bursts are occasionally observed in the metric-decametric (M-D hereon) wavelengths as distinguishable pairs of fundamental (F) and harmonic (H)\footnote{We shall also use $f_F$ and $f_H$ particularly when discussing the emission frequency of F and H.} components. However, they have never been identified as such in the longer hecto-kilometric (H-K hereon) wavelengths \citep[Review of type III harmonic observation difficulties,][]{Dulk2000}. Although attempts have been made to identify the different emission components when the source of the emission was observed in-situ \citep{Kellog80, Reiner19}, such scarce reporting of this fundamental phenomenon is likely due to the lack of appropriate observational capabilities. This has led to a inconsistent understanding the plasma emission mechanism from the Solar corona to interplanetary space.

For the first time, we clearly demonstrate that the most common emission configuration of type III radio bursts in the H-K wavelengths is undoubtedly as fundamental-harmonic pairs. Our findings were made using the observations from the FIELDS instrument suite \citep{BaleFIELDS} on the Parker Solar Probe \citep[PSP;][]{Fox2016} during its close encounters (CE). Furthermore, radio emissions at larger distances (i.e. lower frequencies) are influenced much more by propagation effects \citep[e.g.][]{Krupar20}. This report aims to provide the characteristics of type III radio bursts observed by Parker Solar Probe (PSP) during close encounters when the observer is close to the source and where propagation effects are expected to be less pronounced.



We introduce the experimental details of the study in Section \ref{Sec:observations}, and the spectral characteristics of the type III F-H pairs in Section \ref{Sec:spec_characteristics}. We provide additional evidence for the existence of the F-H pairs through polarization measurements in Section \ref{Sec:polarization}. 



 \begin{figure*}
 \centering
    \includegraphics[width=0.97\textwidth]{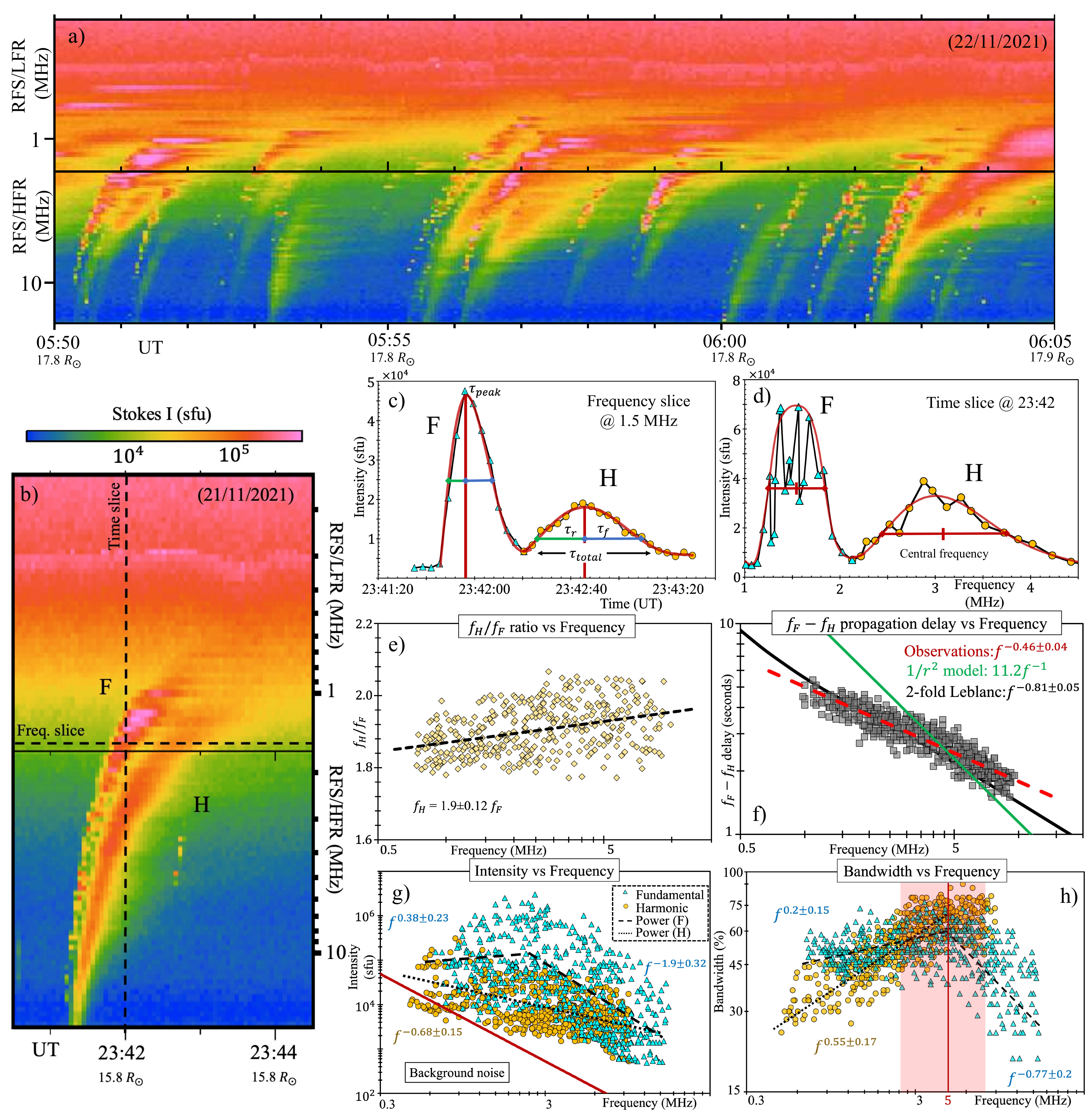}
  \caption{Statistics of spectral characteristics obtained from the analysis of 31 type III F-H pairs. Panel (a) shows a 15-minute interval on 22 November, 2021 where 14 out of 17 type III bursts are F-H pairs. Panel (b) shows an example of a typical type III radio burst with F-H emission and fine structures observed on 21 November, 2021. The dashed lines represent the frequency (1.5~MHz) and time (23:42 UT) slices from the spectrogram which are presented in panels (c) and (d). In panels (a) and (b), the x-axis marks the observer's distance from the Sun. The data points (blue triangles for fundamental and yellow circles for harmonic) are fitted with an exponentially-modified Gaussian (red curve). Here, $\tau_r$ and $\tau_f$ are the rising and falling times, $\tau_{peak}$ is the peak intensity. The statistics of the $f_H/f_F$ ratio, and the $f_F-f_H$ propagation delay as dependencies of frequency are presented in panels (e) and (f). Panels (g), and (h) present the intensity, and bandwidth as a function of the frequency, respectively. F and H are represented by the blue triangle and yellow circle markers. For a one-to-one comparison with F, the results of H are presented as $f_H/2$. The average background noise (QTN + galactic) is illustrated in panel (f) by the red line which scales as $1/f^{2.2\pm0.4}$. The vertical red line and the red shaded region in panel (g) mark the peak bandwidth and its spread.}
  \label{Fig:figure_2}
  \end{figure*}


\section{Experimental details} \label{Sec:observations}

\begin{table}[ht]
    \centering
    \begin{tabular}{c|cc|c}
        Date & Start & Stop & Distance \\
        (dd/mm/yyyy)&(UT)&(UT)&(AU)\\
        \hline \hline
        13/09/2020            & 18:54:40 & 19:08:00 & 0.45 \\\hline
        26/04/2021            & 03:07:00 & 03:09:00 & 0.18 \\\hline
        27/04/2021            & 10:23:51 & 10:26:00 & 0.13 \\ \hline
        02/05/2021            & 17:52:00 & 17:56:00 & 0.18 \\ \hline
        03/05/2021            & 15:36:50 & 15:42:00 & 0.21 \\ \hline
        08/08/2021            & 21:04:10 & 21:07:20 & 0.09 \\ \hline
                                & 16:13:10 & 16:15:30 & 0.16 \\
        18/11/2021             & 18:05:10 & 18:10:00 & 0.15 \\
                           & 20:19:20 & 20:21:40 & 0.15 \\ \hline
        21/11/2021            & 23:41:20 & 23:43:10 & 0.07 \\
            	        & 23:56:00 & 23:57:10 & 0.07 \\ \hline
                           & 02:45:30 & 02:47:10 & 0.08 \\
                           & 03:30:00 & 03:32:15 & 0.08 \\
                           & 03:49:40 & 03:52:20 & 0.08 \\
                           & 04:19:40 & 04:21:30 & 0.08 \\
                           & 04:42:40 & 04:47:30 & 0.08 \\
        22/11/2021         & 06:02:10 & 06:06:00 & 0.08 \\
                           & 06:43:20 & 06:46:00 & 0.08 \\
                           & 07:03:20 & 07:06:00 & 0.08 \\
                           & 10:12:15 & 10:16:00 & 0.08 \\
                           & 11:39:20 & 11:41:10 & 0.09 \\
                           & 12:25:59 & 12:31:00 & 0.09 \\
                           & 23:30:30 & 23:32:00 & 0.11 \\\hline
        23/11/2021         & 08:58:30 & 09:00:30 & 0.13\\
                           & 12:35:25 & 12:36:40 & 0.14 \\\hline 
        24/11/2021         & 11:08:05 & 11:09:30 & 0.18\\
                           & 13:52:10 & 13:54:00 & 0.18 \\ \hline
                           & 01:35:45 & 01:38:30 & 0.23 \\
        26/11/2021         & 07:35:15 & 07:38:30 & 0.24 \\
                           & 20:29:50 & 20:31:00 & 0.26 \\\hline
        27/11/2021         & 07:48:05 & 07:49:50 & 0.28 \\ \hline \hline
    \end{tabular}
    \caption{The list of 31 F-H pairs of type III bursts analysed in this study. Together with the start time and the end time of the burst, we provide also the radial distance of PSP at the time of observation. The level 2 FIELDS/RFS data (intensities are in units of power spectral density, $V^2/Hz$) is open for public and can be accessed from \url{https://fields.ssl.berkeley.edu/data/.} }
    \label{table_bursts}
\end{table}

In this study, we employed the remote sensing measurements made by the radio frequency spectrometer \citep[RFS;][]{Pulupa17}, which combines observations from both the high frequency receiver (HFR) and low frequency receiver (LFR). Both receivers have a $\sim$4\% frequency resolution. Figure \ref{Fig:figure_2}a  presents both the HFR and LFR measurements of a 15-minute interval where a majority (14 out of 17, i.e.$>$80\%) of the type III bursts are F-H pairs. Figure \ref{Fig:figure_2}b presents a typical type III burst with F and H components during the 10th CE of PSP. We analyzed bursts observed during and after the 6th CE due to the enhanced 3.5s temporal resolution. 

We analysed 31 type III radio bursts observed during the 6th--10th CE of PSP, see Table. \ref{table_bursts}. Firstly, we obtained calibrated flux in units of flux density, i.e. W m$^{-2}$ Hz$^{-1}$ or solar flux units (sfu) following the methodology described in \cite{Page22}. Using the effective antenna length ($L_{eff} = L_{34}/L_{12} = 0.99 \pm 0.01$) and the capacitive gain factor ($\Gamma = 0.32$), and the impedence of free space ($Z_0 =$377$\Omega$), in the following relationship from \cite{Pulupa17}, flux density can be estimated,

\begin{equation}
    P[V^2/Hz] = S[W/m^2/Hz] \times Z_0 L_{eff}^2 \Gamma^2
\end{equation} 

To avoid potential convolutions arising from consecutive bursts, we have specifically chosen 31 isolated bursts. Our selection criterion for brightness required each burst to surpass the background level by at least 500 sfu at 15 MHz. Figure \ref{Fig:figure_2}g shows the steep increase of the total background noise (solid red line) below 5 MHz. Here, the level of noise is a pre-event average of the 31 bursts observed between CEs 6 through 10 and it scales close to $f^{-2.2 \pm 0.4}$, where $f$ is the frequency. The presence of a standard deviation in this context is attributed to the variability of the pre-event background, which is dependent on the electron density/temperature during each specific period of the bursts. In combination with the fact that the measurements were conducted during active CEs, this may provide an explanation for the observed shallower scaling law in comparison to the one obtained by \cite{Liu23}, i.e. $f^{-3}$. An important point to clarify is that we did not differentiate between the quasi-thermal noise (QTN) and galactic noise. The galactic noise is frequency-dependent and can reach a maximum of 1000 sfu at close to 3 MHz \citep[][]{Page22}. At these frequencies corresponding to the peak of galactic noise, the type III fluxes are typically several orders of magnitude higher.

\section{Spectral characteristics} \label{Sec:spec_characteristics}

Proximity to the Sun and to the source of the radio emission provided increased sensitivity to unique spectral features of type III radio bursts very rarely or almost never observed in the H-K wavelengths, such as striations and harmonics. These features provide important findings contributing to our understanding of the nature of the plasma emission process at the H-K wavelengths. Although, recent observations have reported fine structures of interplanetary type III bursts \citep{Pulupa20, ChenL21, Jebaraj23}, the observations of type III F-H pairs have yet to be proven conclusively. 

The PSP observations during the CE's show that a significant number of the type III bursts are observed as F-H pairs, regardless of their emission intensity. To emphasize the rate of occurrence, a 15-minute time interval was randomly selected during November 22, 2021. Within this interval, it was discovered that out of the 17 type IIIs, 14 were F-H pairs, accounting for slightly over 80\% (Fig. \ref{Fig:figure_2}a). This finding highlights that the majority of type III bursts observed in this frequency range during CE's are F-H pairs. Similarly, type III radio bursts were visually identified and included in the general statistics for the occurrence rates of F-H pairs during CEs. Only type III bursts that exceeded the background by at least 500 sfu at 15 MHz were taken into account, while bursts occurring in close proximity (more than two type III F-H pairs within a minute) were excluded to prevent signal convolution. It should be noted that large type III storms associated with eruptive events, such as those on 26/04/2021 and 27/04/2021, were not included in the statistical analysis. However, other active periods with relatively lower occurrence rates of type III that still satisfied our criteria were included, such as the type III storm on 22/11/2021. In order to differentiate between occurrence rates during active periods (storm, S) and calmer periods (non-storm, NS), separate statistical analyses were conducted for each. The results of this analysis are presented in Table \ref{table_CE_stats}. These findings provide further quantitative confirmation that a large number of the type III bursts observed during PSP CEs are F-H pairs. It is worth noting that while there was a significant occurrence of type III F-H pairs during the storms (78\%), slightly lower yet still significant rates of occurrence were also observed during calmer periods (70\%).

A customary disclaimer regarding visual identification is that it possesses certain drawbacks, including convolution from multiple type IIIs, which can appear as a single burst. In order to mitigate or substantially reduce potential errors, a rigorous methodology for F-H pair identification has been implemented. Only bursts that met the following three criteria were chosen:

\begin{itemize}
    \item The F-H pairs exhibit a relationship where $f_H$ is approximately twice $f_F$.
    \item The F-H pairs are morphologically distinguishable, with F being structured and H being diffuse and smooth.
    \item The polarization of the F-H pairs is morphologically distinguishable. For more information on the polarization of F-H pairs, please refer to Section \ref{Sec:polarization}.
\end{itemize}

\begin{table}[!ht]
    \centering
    \begin{tabular}{c|cc|c}
        Close & Type III & Type III & Rate \\
        encounter &bursts&F-H pairs&of \\
        CE &&&occurrence \\
        \hline \hline
        CE 6 (NS)            & 4 & 2 & 50\% \\\hline
        CE 7 (NS)           & 28 & 19 & 68\% \\\hline
        CE 8 (S)            & 1167\footnote{The intense type III storm after the CME at 11:00~UT on 26/04/2021 until 16:00~UT on 27/04/2021 was not included in the statistics.} & 812 & 70\% \\
        (NS) &107&71&66\%\\\hline
        CE 9 (NS)          & 49 & 32 & 65\% \\\hline
        CE 10 (S)            & 1877 & 1573 & 84\% \\
        (NS) & 142 & 110 & 77\%\\\hline \hline
        Total (S)                 & 3044 & 2385 & 78\% \\
        (NS) & 330 & 234 & 70\%\\\hline \hline
    \end{tabular}
    \caption{Occurrence rate of type III burst F-H pairs during PSP close encounters (CE) 6 till 10. The occurrence rate during type III storms (S), and quiet periods (NS) are separated.} 
    \label{table_CE_stats}
\end{table}


As presented in (Fig. \ref{Fig:figure_2}a), the F- and H- components of the type III bursts show distinct morphological features thought to be related to the different mode-conversion mechanisms \citep[][]{Ginzburg58, Papadopoulos79, Melrose80, Krasnoselskikh19, Tkachenko21}. Notably, F exhibits a strongly structured spectral appearance, while the H-emission is significantly more ``diffuse" and may occasionally exhibit intensity variations and weak structuring.



The starting frequencies of F- and H-components are different for each burst. Generally, H-component is first seen more often at higher frequencies ($>$15~MHz) whereas F is first seen starting slightly lower. In the example burst shown in Figure \ref{Fig:figure_2}b, the H-component starts at around 19 MHz, while the F-component exhibits fragmentation and is observed first near 15 MHz. The F-component is observed continuing into low frequencies ($<$1 MHz) while H is observed less so at low frequencies. This may partly be due to the domination of QTN close to the local plasma frequency $f_p$ \citep[the spectral tail extends beyond $f_p$ depending on the electron density/temperature, ][]{Zaslavsky11, Liu23}, which ranged between $\sim$ 400 kHz to $\sim$ 1 MHz on average between the first and the tenth CEs. For reference, the $f_p$ at a spacecraft located at 1 AU is $\sim$20 kHz. 

The very high fluxes of the background noise close to the plasma frequency may explain why most type III bursts end around 1 MHz as their signal gets lost in the exceedingly dominant QTN and it's tail. However, since $f_H$ is emitted close to twice the $f_F$, it is expected that the H-component corresponding to the F-component emitted at 10 MHz would be at 20 MHz and therefore is shifted accordingly. This shifting procedure causes some points corresponding to H to be at or in the average background noise. 

In Figure \ref{Fig:figure_2}b, we provide an illustration of a typical type III radio burst, showcasing the primary spectral identification of F-H pairs, and their simultaneous occurrence at $f_p$ and $2f_p$. To study F-H pairs, similar to the one in Figure \ref{Fig:figure_2}b, we analyzed both time and frequency profiles, as depicted in Figures \ref{Fig:figure_2}c and \ref{Fig:figure_2}d. By measuring the difference between the central frequencies of the F and H-components, we obtained the F-H frequency ratio ($f_H/f_F$), which was found to be $f_H$ = 1.9$\pm$0.12$f_F$, close to the theoretically expected $f_H$ = 2$f_F$ (Figure \ref{Fig:figure_2}e). A slight systematic deviation from the predicted $f_H$ = $2f_F$ toward lower frequencies is noticeable from the trend line. Recently, \cite{Melnik18} reported a similar deviation ($f_H$ = 1.87--1.94$f_F$) in the context of metric-decametric bursts. Their findings are in agreement with ours. This deviation is further demonstrated in Figure \ref{Fig:figure_2}f, which displays the measurements of the propagation time-delay between the $f_F$ and $f_H$. In order to obtain these measurements, the temporal deviation of the rising time of the F and H components was compared at any given time for each burst, at the corresponding $f_H/f_F$ ratio. The deviation from the theoretical prediction may stem from the distinct group velocities of radio waves emitted near $f_p$ and those emitted around $\sim 2f_p$. The propagation delay between F and H, based on observations is presented in Figure. \ref{Fig:figure_2}f and it is linearly dependent on frequency as $f^{-0.46\pm 0.04}$ (red dashed line). In Appendix \ref{Sec:propagation delay}, we propose that the physical difference in group velocities could account for a significant proportion of the time delay between the F and H emissions from the source to the observer. The integrals provided in Appendix \ref{Sec:propagation delay} can accommodate any density scaling factor. Figure \ref{Fig:figure_2}f demonstrates a simple $1/r^2$ approximation \citep[depicted by the green solid line;][]{Parker60}, as well as a more advanced 2-fold Leblanc scaling \citep[depicted by the black solid line;][]{Leblanc98}. The delay estimation derived from the 2-fold Leblanc model aligns well with the observations, underscoring its strong dependence on the radial evolution of the electron density profile.

Figure \ref{Fig:figure_2}g presents the peak intensity of both F and H components as a function of frequency. We note that this is the first such comparison of the two emission components in the H-K wavelengths. To make a one-to-one comparison of the F and H intensities, we shifted the points representing H to the frequency of F (i.e. $f_H/2$). The results indicate that the peak intensity of the F-emission increases rapidly towards 5--2 MHz, peaking at $\sim$3~MHz and then slowly declining towards low frequencies ($<$1 MHz). Previous studies \citep[e.g.][]{Weber77} have reported on the radio power peaking close to 1~MHz. A statistical trend of the mean values can be established using a piece-wise fitting with two power-laws \footnote{The piece-wise laws were obtained to be continuous and the limits are chosen based on the transition at which the fit parameters change significantly.}. One with a spectral index, $f^{-1.9\pm0.32}$, at high frequencies (19.2--2 MHz) and a flatter $f^{0.38\pm0.23}$ at low frequencies (2.5--0.75 MHz). Meanwhile, H shows a systematic increase towards low frequency and can be fit with a single power-law with a spectral index of $f^{-0.68\pm0.15}$. It is also evident from the standard deviations that F shows strong intensity variations at all frequencies and therefore presents a large spread of values compared to H. Figure \ref{Fig:figure_2}g demonstrates that the two components have comparable peak intensities at high frequency ($<$10 MHz) and low frequency ($<$1 MHz).

To have an estimate of the physical characteristics of the exciter, we measured the drift rates of F \& H at peak intensity and considered the drift rates of the half-power rising and falling as the standard deviation. We then assumed a 2-fold Leblanc electron density model \citep[][]{Leblanc98} which is often considered for H-K radio bursts \citep[][]{Jebaraj20}. This analysis yielded an average exciter speed of 0.15$c \pm$0.05$c$ for F, and 0.14$c \pm$0.06$c$ for H. The exciter speed of the example burst presented in Figure \ref{Fig:figure_2}b is 0.16$c \pm$0.05$c$ for F, and 0.15$c \pm$0.06$c$ for H. Such speeds are considered nominal type III exciter speeds \citep[][]{Dulk84}. It should be noted that the choice of electron density model introduces an error and therefore this result should be treated only as a first-order approximation. F-H pairs are produced by the same exciter, and as a result, the measured spectral drifts are generally assumed to be identical. 
The difference in spectral drift of about 0.01$c$ could possibly be attributed to a combination of propagation effects (Appendix. \ref{Sec:propagation delay}) and different emission mechanisms.


In addition to the spectral characteristics reported above, we have also analyzed the bandwidth ($df/f$) of the F-H components, and the asymmetry of the type III time profiles ($\tau_f/\tau_r$, where $\tau_f$ is the falling time and $\tau_r$, the rising time). 



\subsection*{Bandwidth}

The bandwidth of the type III burst at any given time instance is calculated as the frequency difference ($df$) between the half-power maximums (Fig. \ref{Fig:figure_2}d). The relative bandwidth is then calculated by dividing $df$ by the central frequency $f$. Since $f_H$ equals 2$f_F$, the H-emission is shifted to $f_F$ for a one-to-one comparison, and the results are presented in Figure \ref{Fig:figure_2}h. The result indicates that the peak relative bandwidth of both F and H are found in a similar range of frequencies. Fig. \ref{Fig:figure_2}h also shows that the relative bandwidth of the H-components systematically decreases with decreasing frequency. This may be attributed to the lower intensity of H and the increasing background noise (Fig. \ref{Fig:figure_2}g), which makes it difficult to accurately measure the true bandwidth using just the half-power maximums. 


Figure \ref{Fig:figure_2}h shows that the relative bandwidth of F can be between 20 and 65\% of their central frequency. Smaller bandwidths (20--40\%) are found close to the starting frequencies (13--8 MHz) after which they grow rapidly towards 6--2.5 MHz, where the average bandwidth is about 45--60\%. The bandwidth does not grow below 2.5 MHz as much and remains close to 50\%. Measuring bandwidth below 1 MHz might induce errors due to the steep increase in background noise (Fig. \ref{Fig:figure_2}g). We perform a piece-wise fitting of the results using two power-laws, obtaining a spectral index of $f^{-0.77\pm0.2}$ in the 19--5 MHz range, and a flat $f^{0.2\pm0.15}$ for the frequencies between 6--0.75 MHz. 

Meanwhile, the bandwidth of H ranged between 30 and 80\% (Fig. \ref{Fig:figure_2}h). The largest bandwidths (60--85\%) were measured at frequencies $\sim$13--5 MHz which is twice the frequency range at which the largest bandwidths of F were found ($\sim$6--2.5 MHz. Due to the constant decrease in the bandwidth, a single power-law with a spectral index of $f^{-0.55\pm0.17}$ fits best. The rapid increase in bandwidth may be due to the divergence of magnetic field lines at heights corresponding to the frequencies between 19 and 5 MHz. 


 \begin{figure*}
  \centering
    \includegraphics[width=1\textwidth]{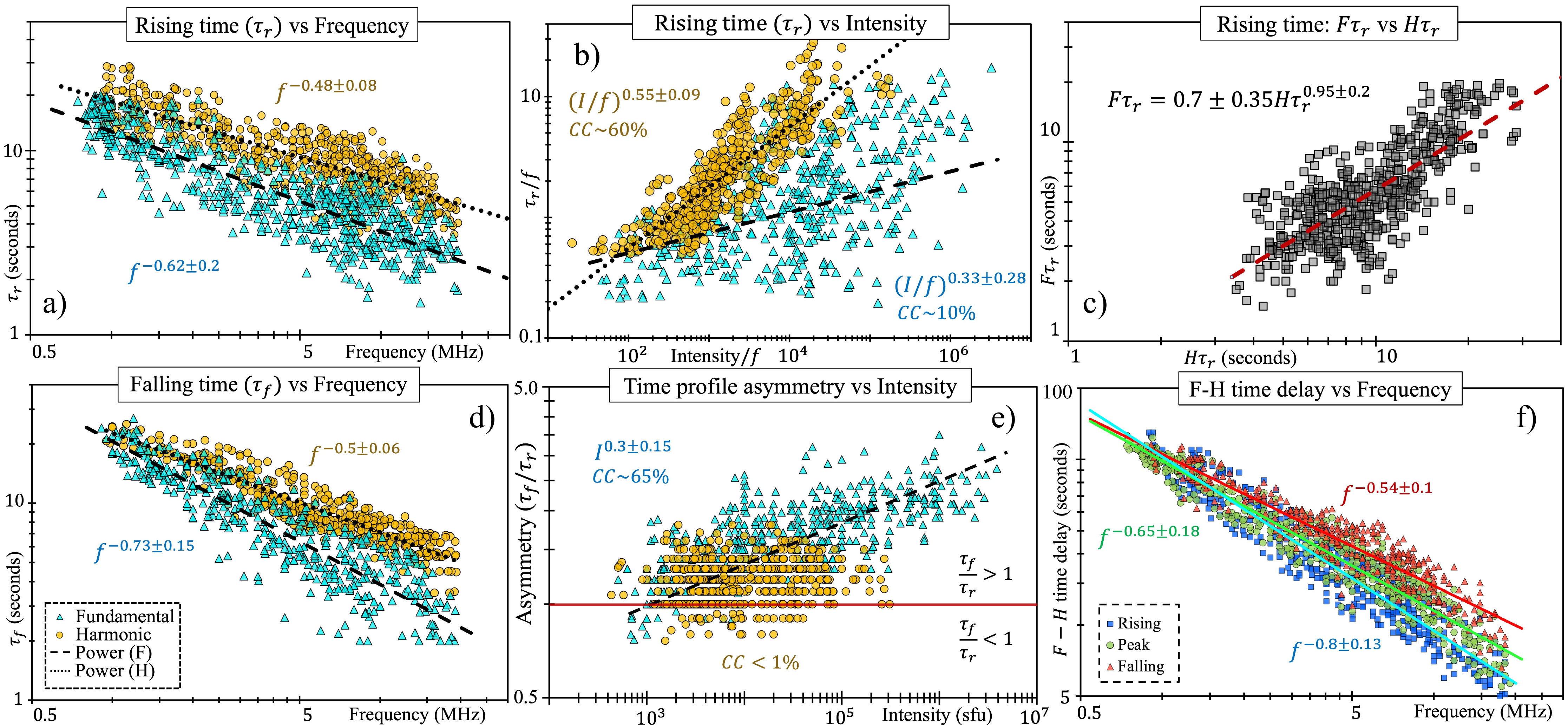}
  \caption{Time-profile characteristics of the 31 type III F-H pairs. The rising time of F (blue triangles) and H (yellow circles) as a function of frequency in panel (a), and as a function of intensity in panel (b). The one-to-one relationship between the rising time of F and H is shown in panel (c). The falling time of F and H as a function of frequency is presented in panel (d), while the time profile asymmetry ($\tau_f/\tau_r$) as a function of intensity is shown in panel (e). The time delay between the F and H-components at the rising and falling phase (at half-power) and at the peak power as a function of frequency is shown in panel (f).}
  \label{Fig:figure_3}
  \end{figure*}

\subsection*{Time profile asymmetry}

The time profiles of type III radio bursts are intrinsically asymmetric \citep[][]{Aubier72, Suzuki85book}. This asymmetry can be used to estimate the two-phase evolution of the beam, namely, the growth of the instability and the damping time scales \citep[][]{Krasnoselskikh19}. Figure \ref{Fig:figure_2}c demonstrates how the two phases can be measured, an exponentially-modified Gaussian fit is used and the peak intensity is marked as $\tau_{peak}$. The fitting procedure for the exponentially-modified Gaussian is described in detail by \citep[][]{Gerekos23} and can also be found in Appendix. \ref{Sec:exgauss}. The values on either side of the $\tau_{peak}$ are the half-power widths, representing the asymmetry of the time-profile which is interpreted as being due to a result of beam-generated Langmuir wave spectrum's evolution \citep[][]{Voshchep15a, Voshchep15b}. In this report, $\tau_f$ is distinguished from $\tau_d$, which is the “exponential” decay time measured at intensities much lower than the half-power \citep[$\sim$ 10\% peak intensity, ][]{Krupar20} and is widely considered when discussing electromagnetic wave diffusion and other propagation effects \citep[e.g.][]{Alvarez73}. 


Figure \ref{Fig:figure_3} presents the different time profile characteristics and their relationship with frequency, intensity, and each other. $\tau_r$ of F and H are shown as functions of frequency in Fig. \ref{Fig:figure_3}a and it is noticeable that the $\tau_r$ of F is considerably faster than that of H. The power-law trends indicate that the $\tau_r$ of F  ($f^{-0.62\pm0.2}$) increases at a slightly faster rate than the $\tau_r$ of H ($f^{-0.46\pm0.08}$) with decreasing frequency. Taking into account the standard deviation of our fits, both F and H scale close to $1/\sqrt{f}$ with frequency. It is worthwhile to note that the large spread in the values of $\tau_r$ of F corresponds to the large variations in the emission intensity. While the mean $\tau_r$ scales slightly over $1/\sqrt{f}$, some individual F-components may trend close to $1/f$. 


Next, we compare the $\tau_r$ of F and H as a function of intensity. The $\tau_r$ of F is primarily due to the increment of the instability and growth of Langmuir waves \citep[][]{Krasnoselskikh19, Jebaraj23}. And such, it is largely dependent upon the characteristics of the beam, and the density fluctuations. However, the $\tau_r$ of H is not as straightforward and is a dominated by the non-linear times associated with the coalescence of the primary and reflected Langmuir wave. The results presented in Fig. \ref{Fig:figure_3}b demonstrates this as $\tau_r$ of F shows a large spread in values (Pearson's correlation coefficient of $\sim$10\%) while $\tau_r$ of H shows a systematic growth with respect to intensity (Pearson's correlation coefficient of $\sim$60\%).

Following this, we have analyzed the direct relationship between $F\tau_r$ and $H\tau_r$, which when considering the mean shows that $F\tau_r$ is $\sim$70\% of the $H\tau_r$. The mean spectral index of $1/f$ further corroborates the almost similar scaling laws obtained for F and H, i.e. $1/\sqrt{f}$ in Fig. \ref{Fig:figure_3}a. 

Measuring the $\tau_r$ of both F and H components at the deca-hectometer wavelengths (19.2 -- 1 MHz) is mostly straightforward, except at the lower frequencies where the $\tau_f$ of F may become convoluted with the $\tau_r$ of H. There are three limiting factors when measuring $\tau_f$, namely, the temporal resolution, increasing background noise, and the expected increase in the $\tau_f$ with decreasing frequency. The first one is a technical issue, while the second one is a consequence of the increased plasma density during CEs, which affects the time profile of both F and H. The final factor is a physical issue arising due to the increase in $\tau_f$ of F becoming increasingly convoluted with the time profile of H even at higher frequencies. Nevertheless, it is still possible to measure the $\tau_f$ (half-power width of the falling time from $\tau_{peak}$) for a limited number of cases. Bearing this in mind, we have measured the $\tau_f$ of F and H whenever possible. We have not measured the “exponential” decay \citep[$\tau_d$,][]{Krupar20} due to the aforementioned reasons which are further enhanced making it difficult to identify the $\tau_d$ of F.

Figure \ref{Fig:figure_3}b shows the results from measuring $\tau_f$ of the F and H components. We find a linear trend with respect to frequency in the case of both F and H. Similar to $\tau_r$, the large standard deviation in the measurements of F is due to the spectral structuring. Meanwhile, the standard deviation for the measurements of H is relatively small. The linear trend is fitted using a power-law with spectral index $f^{-0.5\pm0.06}$ which is similar to $\tau_r$ of H. In the case of F, the fitted power-law has a spectral index of $f^{-0.73\pm0.15}$. Considering only the mean, the scaling law for $\tau_f$ of F can simply be considered to be $1/f^{3/4}$.




 \begin{figure*}
  \centering
    \includegraphics[width=1\textwidth]{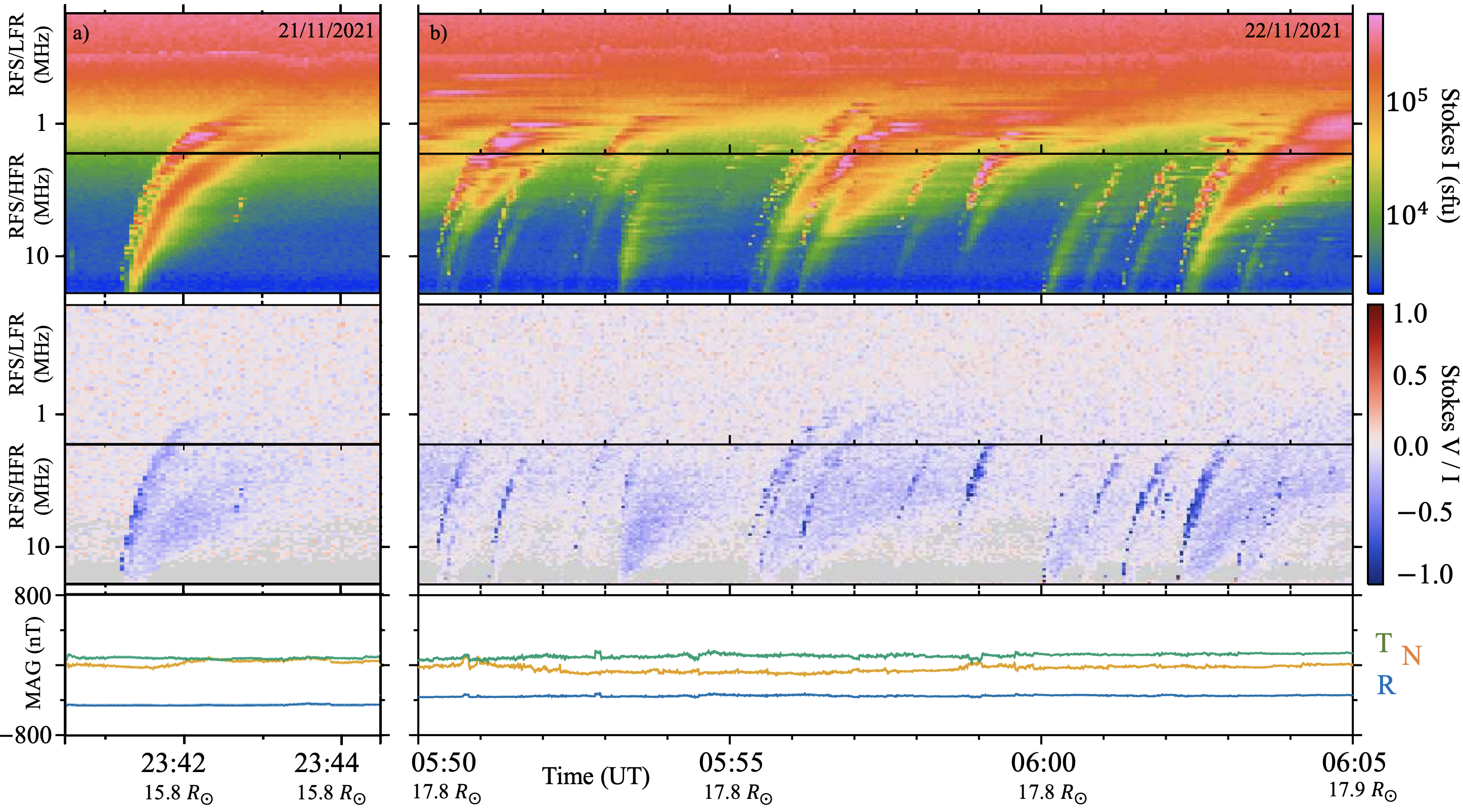}
  \caption{Examples with polarization. The top two panels show the Stokes intensity $I$ parameter from HFR and LFR, for the same example Type III burst shown in Figure \ref{Fig:figure_2} (panel a) and for a longer 15 minute time period on 22/11/2021 (panel b). The next two show the corresponding relative Stokes polarization $V/I$. Separate polarized components are visible for both F and H components of the burst. The bottom panel shows the magnetic field in RTN coordinates, with the field dominated by a negative (sunward) radial component.}
  \label{Fig:figure_polarization}
  \end{figure*}
As demonstrated in Figure \ref{Fig:figure_3}b, the presence of strong intensity variations in the F-emission can result in a drastic spread in rising time. Similarly, the relationship between  $\tau_r$ and $\tau_f$ cannot be fully understood without taking into account the intensity of the emission. Therefore, we investigated the change in $\tau_f$/$\tau_r$ as a function of intensity. Figure \ref{Fig:figure_3}e presents this result and the first thing to note is that $>$80\% of the time-profiles were asymmetric. The sense of asymmetry was where $\tau_f$ was larger than $\tau_r$ (i.e. $\tau_f/\tau_r > 1$). The remaining time profiles were classified into two categories: those where $\tau_r$ was greater than $\tau_f$ (i.e. $\tau_f/\tau_r < 1$), and those where $\tau_r$ and $\tau_f$ were equal, resulting in perfect symmetry (i.e. $\tau_f/\tau_r = 1$).

If we were to compare the symmetry of F- and H- separately, the time-profiles of the F-components were predominantly in the $\tau_f/\tau_r > 1$ regime and were sensitive to the emission intensity. This is demonstrated further by the presence of $\sim \tau_f/\tau_r = 1$ and $\tau_f/\tau_r < 1$ regimes when the intensity was low. Statistically, a linear trend with a Pearson's correlation coefficient of $\sim$65\% was found for the time profiles of F-components. Meanwhile, the results of the H-component presented in Fig. \ref{Fig:figure_3}a indicated no obvious relationship between the intensity and the symmetry (Pearson's correlation coefficient $<$1\%). The H-component was also likely to be far more symmetric ($\tau_f/\tau_r = 1$) in comparison to the F-component. In terms of asymmetry, we noted only minor asymmetry which was irrespective of the emission intensity.  

This result was different from the one obtained in Figure \ref{Fig:figure_3}b, where no significant correlation was found between the emission intensity and the rising time. When considering the falling time as well, the asymmetry of the F-emission was correlated strongly with the variations in intensity.

Additionally, in Fig. \ref{Fig:figure_3}f, we report for the first time the time delay between the F-H pairs at discrete frequencies. As a result of emissions from distinct regions, F and H-emissions emitted at a specific frequency are anticipated to become more divergent as the frequency decreases. This divergence offers a potential means of measuring the speed of the exciter. This separation between F-H pairs has not been reported previously due to the poor temporal resolution of H-K observations. In Fig. \ref{Fig:figure_3}f, we demonstrate that it is possible to measure the time-delay between the rising (blue squares), peak (green diamonds), and the falling (red triangles) of the frequency-time profile of the F-H pairs. As their emission regions ($f_p$ and $2f_p$) become increasingly separated with decreasing frequency, so too does the time difference between different parts of the F-H pairs. The results indicate that there is a systematically increasing delay between them, which can be best-fitted using power-laws with spectral indices; $f^{-0.8\pm0.13}$ (rising), $f^{-0.65\pm0.18}$ (peak), and $f^{-0.54\pm0.1}$ (falling). The time difference between the rising and falling phases is measured by taking the half-power maximums of the F-H pairs (shown in Fig. \ref{Fig:figure_2}b). By utilizing a simple density model \citep[2-fold,][]{Leblanc98}, we derived 0.2$c (\pm0.026c)$, 0.16$c (\pm0.03c)$, and 0.11$c (\pm0.018c)$ for the rising, peak, and falling times, respectively (Fig. \ref{Fig:figure_3}f). This finding supports the estimate we presented in Section \ref{Sec:spec_characteristics}. 

\section{Polarization} \label{Sec:polarization}

When the fundamental and harmonic components of a radio burst are clearly distinguishable, we can also study the polarization properties of the individual components. While a detailed description of the polarization properties is beyond the scope of this letter, we can make a few initial observations, using the same examples presented in Figure \ref{Fig:figure_2}a \& b. For these example bursts, we calculate the Stokes parameters as in \citet{Pulupa20}. The polarization measurements provide additional confirmation that the observed bursts are indeed F-H pairs. 

For F-H Type IIIs where polarization is evident, its properties are broadly consistent with those described in \citet{DulkSuzuki1980} from ground-based observations at 24-220 MHz.

The degree of polarization (DOP), represented using the ratio of Stokes $V$ to $I$ parameters, is strongest near the leading edge of the F component, and significantly weaker for the H component. In the example in Figure \ref{Fig:figure_polarization}a, the F Stokes $V/I$ reaches maximum of $\sim$0.7, while $V/I$ for the H component is at 0.2--0.3. As is the case with the intensity, the circular polarization for the F component is highly variable, while the H component is smoother. Figure \ref{Fig:figure_polarization}b presents a longer, 15 minute duration during the same CE on 22/11/2021. A number of type III F-H pairs are distinguished and the Stokes $V/I$ of F reaches or exceeds $\sim$0.5 for all cases with some even reaching maximum close to $\sim$1.0. As with the example presented in Figure \ref{Fig:figure_polarization}a, the H-component shows substantially lower DOP $\sim$ 0.3 on average. The F-component also exhibits strong variations while the H is diffuse. It is also worthwhile to note that the polarization is substantially higher for fine-structures within F-emission. 

The high DOP in the measurements is attributed to the angle between the magnetic field at the emission source and the direction to the spacecraft. Consequently, it is not surprising that the DOP varies across different source regions, as demonstrated by the groups of type III bursts examined in a recent study \citep{Dresing23}. Another study by \cite{Pulupa20} analyzed a type III radio burst storm during the PSP's second CE and similarly found a high DOP. Additionally, prior studies based on observations at 1 AU \citep{Reiner07} reported much smaller DOP values and were unable to distinguish between F and H. The polarization measurements presented here unequivocally indicate that the observed bursts are F-H pairs.


%

The sense of the circularly polarized emission is always the same between the F and H component, and is determined by the direction of the magnetic field at the source region. For the F component, which is emitted near the plasma frequency $f_p$, the $x$-mode radiation produced at the source region has a frequency below $f_p$ and cannot propagate to the observer. Therefore the observed emission should be in the $o$-mode, which is left-hand circularly polarized (LHC) when the radial component of the source region magnetic field $B_r > 0$, and right-hand (RHC) when $B_r < 0$. Although the sense of polarized F emission is determined by the direction of the field, it is not well understood what controls the degree of polarization, i.e., why emission which is restricted to one mode is not 100\% polarized. Reflection off of regions with enhanced density \citep{Melrose2006} can result in depolarization, and simulations which include effects of density variation \citep{Kim2007,Kim2008} indicate that it is possible to produce emission in $x$ and $o$-modes simultaneously. Such a scenario may also explain the high polarization of the fine-structures within F-emission where density inhomogenities are relatively low \citep[][]{Jebaraj23}. The general decrease in polarization observed below 1.5~MHz may be attributed to plasma inhomogenities between the source and the observer, the aforementioned mode coupling between $x$- and $o$-modes, and directivity of emission with respect to the observer. For the H component, the sense of polarization matches that of the F component. The degree of H polarization is related to the ratio of the cyclotron frequency $f_c$ to $f_p$ in the source region \citep{DulkSuzuki1980}. 

The proximity of PSP to the source of the emission and the fact that the magnetic field trends more radial in the inner heliosphere \citep{Badman2021} allows us to directly compare the sense of polarization using the in situ magnetic field data. In Figure \ref{Fig:figure_polarization}, we can confirm that the RHC sense of polarization for F and H matches the negative sign of $B_r$.




\section{Conclusions} \label{Sec:conclusions}

In this letter, we have reported for the first time clearly distinguishable fundamental-harmonic pairs among type III radio bursts observed in the hecto-kilometric wavelengths. We attribute this finding to the close proximity of the observer to the source and the enhanced time and frequency resolution of the FIELDS/RFS receivers onboard the Parker Solar Probe. The main findings of this letter are listed below:

\begin{enumerate}
    \item {We found that a majority (more than seventy percent) of the type III radio bursts observed during PSP CEs 6--10 are F-H pairs. We also found that their occurrence rate is slightly higher during storm periods.}

    \item {The morphology of the F-emission exhibits strong structuring, while the H-emission appears diffuse in both Stokes $I$ and $V$, resembling the type IIIb-type III pairs observed in metric-decametric wavelengths.}

    \item {There is a systematic delay in propagation between the F and H-emission which increases with decreasing frequency and offsets the theoretically expected $f_H = 2 f_F$ to $f_H = 1.9\pm0.12f_F$. We have demonstrated that this is due to the difference in group velocity of the F and H.}

   \item {We found that the time-profile asymmetry of F is well correlated to the intensity of the emission. However, the H-emission shows no such correlation.}
   
    \item {Our results indicate that the rising time of F is consistently faster than the H regardless of intensity. The variations of rising time of F-emission is also strongly associated with its intensity i.e., the more intense the emission, faster the rise.}

    \item {The F-emission is highly polarized ($\sim$ 50\% on average in the 19--1.5MHz frequency range) with some bursts showing $>$90\% polarization. Meanwhile, the H-emission is weakly polarized ($\sim$ 30\% or lower on average). The high polarization of F indicates that it is generated predominantly as $o$-mode radiation, while H is a mix of both $x$- and $o$-mode radiation.}

    \item {The duration parameters ($\tau_r$ and $\tau_f$) of the harmonic scale linearly with frequency at a rate of $1/\sqrt{f}$, while that of the fundamental exhibits complexity due to strong intensity variations. Specifically, the $\tau_r$ scales close to $1/\sqrt{f}$, while $\tau_f$ scales as $1/f^{3/4}$.}

\end{enumerate}

The observational evidence and statistical results presented in this report on the observational characteristics of the fundamental and harmonic emission components near the source offer new avenues for exploration in the hecto-kilometric wavelengths. Observing radio bursts in close proximity to their source allows us to deepen our understanding of fundamental plasma processes and the generation of radio emission in inhomogeneous plasma.
Aside from the fundamental aspects of the beam-plasma system, our study builds upon previous research conducted by \cite{Krupar20} and offers valuable enhancements for probing the evolution of density turbulence in the coronal and solar wind plasma.
The impact of multi-vantage point radio observations on the analysis of solar energetic particle (SEP) transport continues to grow rapidly. Type III bursts serve as a powerful tool for comprehending the propagation path and plasma conditions \citep[e.g.,][]{Dresing23, Jebaraj23b}. Consequently, distinguishing between F and H aids in refining techniques used to pinpoint the source of type III bursts and improve source propagation estimation \citep[direction finding,][]{Lecacheux78}.
Furthermore, this discovery greatly contributes to the longstanding challenge of distinguishing radio wave propagation \citep[][and references therein]{Arzner99}. Specifically, the exponential decay provides pivotal insights into the processes influencing radio waves. Due to the lack of radio imaging in the H-K wavelength range, distinguishing the characteristics of the F-H time profiles is crucial.

In forthcoming publications, we will also progress the state-of-the-art by addressing the generation of both the fundamental and harmonic emission which has been predicted by the probabilistic model of beam-plasma interactions \citep[][]{Tkachenko21, Krafft22b} using experimental data from PSP.
Finally, coordinated observations made from the Parker Solar Probe during its close encounters, in conjunction with the Radio Plasma Waves \citep[RPW;][]{Maksimovic20} instrument on board the Solar Orbiter \citep[SolO; ][]{Muller13} may offer additional opportunities to understand multi-scale plasma processes in the high corona and interplanetary space. A likely avenue for future exploration is a survey of type III radio bursts using PSP and SolO for which local Langmuir waves are observed. A similar approach to \cite{Reiner19} would make it possible to distinguish F-H pairs at lower frequencies (i.e. $<$400 kHz). At such frequencies, propagation effects and the effect of a larger spatial extent of the source may result in emission from a wider range of frequencies simultaneously making the peak-frequencies less pronounced.





\section*{acknowledgements}

This research was supported by the International Space Science Institute (ISSI) in Bern, through ISSI International Team project No.557, “Beam-Plasma Interaction in the Solar Wind and the Generation of Type III Radio Bursts”. 
Parker Solar Probe was designed, built, and is now operated by the Johns Hopkins Applied Physics Laboratory as part of NASA’s Living with a Star (LWS) program (contract NNN06AA01C). Support from the LWS management and technical team has played a critical role in the success of the Parker Solar Probe mission.
V.K., acknowledges financial support from CNES through grants, ``Parker Solar Probe", and ``Solar Orbiter", and by NASA grant 80NSSC20K0697.
J.M., acknowledges funding by the  BRAIN-be project SWiM (Solar Wind Modeling with EUHFORIA for the new heliospheric missions).

\appendix

\section{Exponentially-modified Gaussian} \label{Sec:exgauss}

The type III time profiles which exhibit a rapid Gaussian-like rise, and an exponential decay were fitted using a five-parameter function known as the exponentially-modified Gaussian function \citep[][]{Eli72}. This function, which was also used in \cite{Gerekos23} is expressed as:

\begin{equation} \label{exgaussian}
    g(a,b,\mu,\sigma,\lambda; x) = a \frac \lambda 2 \text{exp}\left\{ \frac \lambda 2 \left( 2\mu + \lambda\sigma^2 - 2x\right) \right\} \left( 1- \text{erf}\left\{ \frac{\mu+\lambda\sigma^2 - x}{\sqrt{2}\sigma}\right\}\right) + b,
\end{equation}

Here, the parameters have specific roles. The parameter $a$ scales the overall magnitude of the burst, while $b$ represents its base level, i.e. pre-event background. The mean ($\mu$) and variance ($\sigma^2$) determine the Gaussian portion of the function. Lastly, the decay rate ($\lambda$) controls the exponential decay part. The error function ($\text{erf}(z)$) is defined as $2(\pi)^{-1/2} \int_0^z e^{-t^2}\text{d}t$. The values on either side of the $\tau_{peak}$ are the half-power widths, representing the asymmetry of the time-profile, i.e. the rising time ($\tau_r$), and falling time ($\tau_f$).

We perform separate fits for both fundamental and harmonic bursts, and then combine them at the intersection point where the goodness-of-fit typically deviates from the $>$95\% level. Time profiles that had a goodness below the optimal 95\% level prior to the intersection are discarded. Additionally, time profiles where the intersection occurs at or before the half-maximum level are also discarded. For the harmonic bursts, we assume the pre-event background or base level to be the same as that of the fundamental bursts (F). As a result, the parameter $b$ remains constant for both F and H at each frequency.

\section{Propagation delay between fundamental and harmonic emission} \label{Sec:propagation delay}


The fundamental and harmonic emission at any given moment are emitted at different frequencies, and therefore propagate with different group velocities. This physical effect has been evaluated analytically in this section.

Let us evaluate the ``time of flight" of the EM wave between source region where the generation occurs to observer.

For this purpose, we use the Hamiltonian description:

\begin{equation}
    \frac{d\mathbf{k}}{dt}=-\nabla \omega 
\end{equation}

\begin{equation}
    \frac{d\mathbf{r}}{dt}=\mathbf{V}_{gr}=\frac{\partial \omega }{\partial 
\mathbf{k}} 
\end{equation}

Here, $V_{gr}$ is the group velocity, and $\omega$ is frequency in radians s$^{-1}$ and is related to $f$ as $\frac{\omega}{2\pi}$.

Let us formulate the problem in simplified version, the wave is generated at some source point, that we shall notify by index $S$ and propagates along the radius and the plasma density depends on the radial distance only. Let the radial dependence be described by an expression,

\begin{equation}
n(R)=n_{S}\left(\frac{R_{S}}{R}\right)^{\alpha } 
\end{equation}

\begin{equation*}
\frac{\omega_{p}(R)}{\omega_{pS}}=\left(\frac{R_{S}}{R}\right)^{\alpha /2}, R=R_{S}\left(\frac{\omega_{pS}}{\omega_{p}(R)}\right)^{2/\alpha } 
\end{equation*}

Then a simple calculation allows one to obtain the following set of equations:

\begin{equation}
    \frac{d\mathbf{k}}{dt}=-\nabla \omega_{p} 
\end{equation}

\begin{equation*}
\mathbf{V}_{gr}=\omega_{p}\frac{\mathbf{k}c^{2}}{\omega_{p}^{2}\sqrt{1+\frac{k^{2}c^{2}}{\omega_{p}^{2}}}}=\frac{\mathbf{k}c}{\omega}c 
\end{equation*}

\begin{equation*}
\frac{d\mathbf{V}_{gr}}{dt}=-\frac{c^{2}}{\omega}\nabla \omega_{p} 
\end{equation*}

\begin{equation*}
\frac{dV_{gr}}{dt}=\frac{\alpha c^{2}R_{S}^{\alpha }}{\omega R^{\alpha +1}}\omega_{p_{S}} 
\end{equation*}

\begin{equation}
\frac{dR}{dt}=\mathbf{V}_{gr} 
\end{equation}

\begin{equation*}
\frac{d^{2}R}{dt^{2}}=\frac{\alpha c^{2}R_{S}^{\alpha }}{\omega R^{\alpha +1}}\omega_{p_{S}}   
\end{equation*}

\begin{equation*}
\frac{dR}{dt}\frac{d^{2}R}{dt^{2}}=\frac{dR}{dt}\frac{\alpha c^{2}R_{S}^{\alpha }}{\omega R^{\alpha +1}}\omega_{p_{S}} 
\end{equation*}

\begin{equation*}
V_{gr}^{2}-V_{grS}^{2}=\frac{2c^{2}}{\omega}\omega_{p_{S}}\left[1-\left(\frac{R_{S}}{R}\right)^{\alpha }\right]   
\end{equation*}

\begin{equation}
V_{gr}^{2}=\left\lbrace V_{grS}^{2}+\frac{2c^{2}}{\omega}\omega_{p_{S}}\left[1-\left(\frac{R_{S}}{R}\right)^{\alpha }\right]\right\rbrace     
\end{equation}

\noindent Knowing the group velocity allows us to evaluate the ``time of flight":

\begin{equation*}
T = \bigints_{R_{S}}^{R}\frac{dR}{\left[V_{grS}^{2}+\frac{2c^{2}}{\omega }\omega_{p{S}}\left(1-\left(\frac{R_{S}}{R}\right)^{\alpha}\right)^{1/2}\right]}
= R_{S}\bigints_{1}^{R/R_{S}}\frac{y^{\alpha /2}dy}{\left[\left(V_{grS}^{2}+\frac{2c^{2}}{\omega }\omega_{p{S}}\right)y^{\alpha }-\frac{2c^{2}}{\omega}\omega_{p{S}}\right]^{1/2}} 
\end{equation*}
\begin{equation}
T = \left(\frac{\omega}{2\omega_{p{S}}}\right)^{1/2}\frac{R_{S}}{c}\bigints_{1}^{R/R_{S}}\frac{y^{\alpha/2}dy}{\left[\left(\frac{\omega V_{grS}^{2}}{2c^{2}\omega_{p{S}}}+1\right)y^{\alpha }-1\right]^{1/2}}\
= \left(\frac{\omega}{2\omega_{p{S}}}\right)^{1/2}\frac{R_{S}}{c}\bigints_{1}^{R/R_{S}}\frac{y^{\alpha /2}dy}{\left[Qy^{\alpha }-1\right]^{1/2}}
\end{equation}

here, $Q=\left(\frac{\omega V_{grS}^{2}}{2\omega_{p_{S}}c^{2}}+1\right)$, and $y=\frac{R}{R_{S}}$. This time can be evaluated for a wave generated at fundamental frequency and for its harmonic. Let us first evaluate the parameter $Q$ for the fundamental frequency.

In order to do that let us analyze the relations between frequencies and k-vectors. For the fundamental frequency the Langmuir wave frequency is written as 

\begin{equation}
\omega_{L}=\omega_{p}\left(1+\frac{3}{2}k_{L}^{2}\lambda_{D}^{2}\right)=\omega_{tF}=\omega_{p}\left(1+\frac{1}{2}\frac{k_{t}^{2}c^{2}}{\omega_{p}^{2}}\right)
\end{equation}

Here, the subscript $t$ indicates a transverse wave (electromagnetic wave) generated with the same frequency as the primary Langmuir wave. So, $k_L$ is the k-vector of primarily generated Langmuir wave, which is given by $k_{L}=\frac{\omega_{p}}{V_{b}}$, and then, $k_{tF}=\sqrt{3}k_{L}\frac{\lambda_{D}\omega_{p}}{c}=\sqrt{3}\frac{\omega_{p}}{V_{b}}\frac{v_{T}}{c}$ is the vector of electromagnetic wave. Using this vector, we can find the group velocity for the fundamental frequency as:

\begin{equation}
V_{grSF}=\frac{k_{tF}c^{2}}{\omega_{p}\left(1+\frac{k_{tF}^{2}c^{2}}{\omega_{p}^{2}}\right)^{1/2}}=\frac{k_{tF}c}{\omega }c=\sqrt{3}\frac{v_{T}}{V_{b}}c
\end{equation}

\noindent Therefore, the parameter $Q$ for fundamental emission is equal to

\begin{equation}
Q_{F}=\left(\frac{3v_{T}^{2}}{2V_{b}^{2}}+1\right)
\end{equation}

And the "time of flight" for the wave generated at fundamental frequency under condition that it propagates radially from the source to the observer without small-angle scattering is evaluated to be:

\begin{equation}
T_{F}=\frac{R_{S}}{\alpha c}\frac{\sqrt{2}}{2}\bigints_{1}^{R/R_{S}}%
\frac{y^{\alpha /2}dy}{(Q_{F}y^{\alpha }-1)^{1/2}}=\frac{R}{\alpha c}\left(\frac{\omega_{pL}}{\omega }\right)^{2/\alpha }\frac{\sqrt{2}}{2}\bigints_{1}^{R/R_{S}}\frac{y^{\alpha /2}dy}{(Q_{F}y^{\alpha }-1)^{1/2}}
\end{equation}

Here, $\omega_{pL}$ is the local plasma frequency. It is easy to make similar calculations for the harmonic emission. Initial k-vector of wave generated by nonlinear wave-wave interaction satisfies the following:

\begin{equation}
2\omega_{p}=\sqrt{k^{2}c^{2}+\omega_{p}^{2}}   
\end{equation}

\begin{equation}
k_{SH}=\sqrt{3}\frac{\omega}{2c} 
\end{equation}

\begin{equation}
V_{grSH}=\frac{\sqrt{3}}{2}c   
\end{equation}

The "time of flight" for the harmonic emission can then be given as,

\begin{equation*}
T_{H}=\bigints_{R_{S}}^{R}\frac{dR}{\left[V_{grSH}^{2}+\frac{2c^{2}}{\omega}\omega_{p_{S}}\left( 1-\left( \frac{R_{S}}{R}\right)^{\alpha }\right) ^{1/2}\right]}
\end{equation*}

\begin{equation*}
T_{H}=\bigints_{R_{S}}^{R}\frac{dR}{\left[ \frac{3}{4}c^{2}+c^{2}\left( 1-\left( \frac{R_{S}}{R}\right) ^{\alpha }\right) ^{1/2}\right]}=\frac{R_{S}}{c}\bigints_{1}^{R/R_{S}}\frac{d(R/R_{S})}{\left[ \frac{7}{4}-\left( \frac{R_{S}}{R}\right) ^{\alpha }\right] ^{1/2}}
\end{equation*}

Changing variable $(R/R_{S})=y$

\begin{equation}
T_H=\frac{R_{S}}{c}\bigints_{1}^{R/R_{S}}\frac{y^{\alpha /2}dy}{\left[\frac{7}{4}y^{\alpha }-1\right] ^{1/2}}=\frac{2R}{\sqrt{7}c}\left( \frac{\omega_{pL}}{\omega}\right) ^{2/\alpha }\bigints_{1}^{R/R_{S}}\frac{y^{\alpha /2}dy}{\left( y^{\alpha }-\frac{4}{7}\right) ^{1/2}}
\end{equation}

The difference of arrival times is presented by the following expression:

\begin{equation}
\Delta T=T_{F}-T_{H}.
\end{equation}

The integrals $T_F$ and $T_H$ can be evaluated numerically for sophisticated electron density profiles. However, as an exercise, a simple calculation for the case $\alpha =2$ is provided here. The time difference ($\Delta T$) may then also be presented in the form of analytic expressions. Under assumption $R>>R_{S}$ one can find the following approximate expressions:

\begin{equation*}
T_{F}=\frac{R}{c\sqrt{2}}\left\lbrace \left( 1-\frac{R_{S}}{R}\right) ^{1/2}+\frac{R_{S}}{R}\ln \left[ \left( \frac{R}{R_{S}}\right)^{1/2}+\left( \frac{R}{R_{S}}-1\right) ^{1/2}\right] \right\rbrace ,
\end{equation*}

and

\begin{equation*}
T_{H}=\frac{2R}{c\sqrt{7}}\left[ 1+\frac{4}{7}\frac{R_{S}}{R}\ln \left( \frac{7R}{4R_{S}}\right) \right] .
\end{equation*}

Rough evaluation gives the following.

\begin{equation}
\Delta T\approx 0.1\frac{R}{c}.
\end{equation}

\bibliography{bibTeX_11_11_2022}{}

\begin{thebibliography}{}
\expandafter\ifx\csname natexlab\endcsname\relax\def\natexlab#1{#1}\fi
\providecommand{\url}[1]{\href{#1}{#1}}
\providecommand{\dodoi}[1]{doi:~\href{http://doi.org/#1}{\nolinkurl{#1}}}
\providecommand{\doeprint}[1]{\href{http://ascl.net/#1}{\nolinkurl{http://ascl.net/#1}}}
\providecommand{\doarXiv}[1]{\href{https://arxiv.org/abs/#1}{\nolinkurl{https://arxiv.org/abs/#1}}}

\bibitem[{{Alvarez} \& {Haddock}(1973)}]{Alvarez73}
{Alvarez}, H., \& {Haddock}, F.~T. 1973, \solphys, 30, 175, \dodoi{10.1007/BF00156186}

\bibitem[{{Aubier} \& {Boischot}(1972)}]{Aubier72}
{Aubier}, M., \& {Boischot}, A. 1972, \aap, 19, 343

\bibitem[{{Badman} {et~al.}(2021){Badman}, {Bale}, {Rouillard}, {Bowen}, {Bonnell}, {Goetz}, {Harvey}, {MacDowall}, {Malaspina}, \& {Pulupa}}]{Badman2021}
{Badman}, S.~T., {Bale}, S.~D., {Rouillard}, A.~P., {et~al.} 2021, \aap, 650, A18, \dodoi{10.1051/0004-6361/202039407}

\bibitem[{{Bale} {et~al.}(2016){Bale}, {Goetz}, {Harvey}, {Turin}, {Bonnell}, {Dudok de Wit}, {Ergun}, {MacDowall}, {Pulupa}, {Andre}, {Bolton}, {Bougeret}, {Bowen}, {Burgess}, {Cattell}, {Chandran}, {Chaston}, {Chen}, {Choi}, {Connerney}, {Cranmer}, {Diaz-Aguado}, {Donakowski}, {Drake}, {Farrell}, {Fergeau}, {Fermin}, {Fischer}, {Fox}, {Glaser}, {Goldstein}, {Gordon}, {Hanson}, {Harris}, {Hayes}, {Hinze}, {Hollweg}, {Horbury}, {Howard}, {Hoxie}, {Jannet}, {Karlsson}, {Kasper}, {Kellogg}, {Kien}, {Klimchuk}, {Krasnoselskikh}, {Krucker}, {Lynch}, {Maksimovic}, {Malaspina}, {Marker}, {Martin}, {Martinez-Oliveros}, {McCauley}, {McComas}, {McDonald}, {Meyer-Vernet}, {Moncuquet}, {Monson}, {Mozer}, {Murphy}, {Odom}, {Oliverson}, {Olson}, {Parker}, {Pankow}, {Phan}, {Quataert}, {Quinn}, {Ruplin}, {Salem}, {Seitz}, {Sheppard}, {Siy}, {Stevens}, {Summers}, {Szabo}, {Timofeeva}, {Vaivads}, {Velli}, {Yehle}, {Werthimer}, \& {Wygant}}]{BaleFIELDS}
{Bale}, S.~D., {Goetz}, K., {Harvey}, P.~R., {et~al.} 2016, \ssr, 204, 49, \dodoi{10.1007/s11214-016-0244-5}

\bibitem[{{Chen} {et~al.}(2021){Chen}, {Ma}, {Wu}, {Zhao}, {Tang}, \& {Bale}}]{ChenL21}
{Chen}, L., {Ma}, B., {Wu}, D., {et~al.} 2021, \apjl, 915, L22, \dodoi{10.3847/2041-8213/ac0b43}

\bibitem[{{Dresing} {et~al.}(2023){Dresing}, {Rodr{\'\i}guez-Garc{\'\i}a}, {Jebaraj}, {Warmuth}, {Wallace}, {Balmaceda}, {Podladchikova}, {Strauss}, {Kouloumvakos}, {Palmroos}, {Krupar}, {Gieseler}, {Xu}, {Mitchell}, {Cohen}, {de Nolfo}, {Palmerio}, {Carcaboso}, {Kilpua}, {Trotta}, {Auster}, {Asvestari}, {da Silva}, {Dr{\"o}ge}, {Getachew}, {G{\'o}mez-Herrero}, {Grande}, {Heyner}, {Holmstr{\"o}m}, {Huovelin}, {Kartavykh}, {Laurenza}, {Lee}, {Mason}, {Maksimovic}, {Mieth}, {Murakami}, {Oleynik}, {Pinto}, {Pulupa}, {Richter}, {Rodr{\'\i}guez-Pacheco}, {S{\'a}nchez-Cano}, {Schuller}, {Ueno}, {Vainio}, {Vecchio}, {Veronig}, \& {Wijsen}}]{Dresing23}
{Dresing}, N., {Rodr{\'\i}guez-Garc{\'\i}a}, L., {Jebaraj}, I.~C., {et~al.} 2023, \aap, \dodoi{10.1051/0004-6361/202345938}

\bibitem[{{Dulk}(2000)}]{Dulk2000}
{Dulk}, G.~A. 2000, Geophysical Monograph Series, 119, 115, \dodoi{10.1029/GM119p0115}

\bibitem[{{Dulk} {et~al.}(1984){Dulk}, {Steinberg}, \& {Hoang}}]{Dulk84}
{Dulk}, G.~A., {Steinberg}, J.~L., \& {Hoang}, S. 1984, \aap, 141, 30

\bibitem[{{Dulk} \& {Suzuki}(1980)}]{DulkSuzuki1980}
{Dulk}, G.~A., \& {Suzuki}, S. 1980, \aap, 88, 203

\bibitem[{{Fox} {et~al.}(2016){Fox}, {Velli}, {Bale}, {Decker}, {Driesman}, {Howard}, {Kasper}, {Kinnison}, {Kusterer}, {Lario}, {Lockwood}, {McComas}, {Raouafi}, \& {Szabo}}]{Fox2016}
{Fox}, N.~J., {Velli}, M.~C., {Bale}, S.~D., {et~al.} 2016, \ssr, 204, 7, \dodoi{10.1007/s11214-015-0211-6}

\bibitem[{{Gerekos} {et~al.}(2023){Gerekos}, {Steinbr{\"u}gge}, {Jebaraj}, {Casillas}, {Donini}, {S{\'a}nchez-Cano}, {Lester}, {Magdaleni{\'c}}, {Peters}, {Romero-Wolf}, \& {Blankenship}}]{Gerekos23}
{Gerekos}, C., {Steinbr{\"u}gge}, G., {Jebaraj}, I., {et~al.} 2023, arXiv e-prints, arXiv:2307.01747, \dodoi{10.48550/arXiv.2307.01747}

\bibitem[{{Ginzburg} \& {Zhelezniakov}(1958)}]{Ginzburg58}
{Ginzburg}, V.~L., \& {Zhelezniakov}, V.~V. 1958, \sovast, 2, 653

\bibitem[{Grushka(1972)}]{Eli72}
Grushka, E. 1972, Analytical Chemistry, 44, 1733, \dodoi{10.1021/ac60319a011}

\bibitem[{{Jebaraj} {et~al.}(2023{\natexlab{a}}){Jebaraj}, {Magdalenic}, {Krasnoselskikh}, {Krupar}, \& {Poedts}}]{Jebaraj23}
{Jebaraj}, I.~C., {Magdalenic}, J., {Krasnoselskikh}, V., {Krupar}, V., \& {Poedts}, S. 2023{\natexlab{a}}, \aap, 670, A20, \dodoi{10.1051/0004-6361/202243494}

\bibitem[{{Jebaraj} {et~al.}(2020){Jebaraj}, {Magdalenic}, {Podladchikova}, {Scolini}, {Pomoell}, {Veronig}, {Dissauer}, {Krupar}, {Kilpua}, \& {Poedts}}]{Jebaraj20}
{Jebaraj}, I.~C., {Magdalenic}, J., {Podladchikova}, T., {et~al.} 2020, \aap, 639, \dodoi{10.1051/0004-6361/201937273}

\bibitem[{{Jebaraj} {et~al.}(2023{\natexlab{b}}){Jebaraj}, {Kouloumvakos}, {Dresing}, {Warmuth}, {Wijsen}, {Palmroos}, {Gieseler}, {Marmyleva}, {Vainio}, {Krupar}, {Wiegelmann}, {Magdalenic}, {Schuller}, {Battaglia}, \& {Fedeli}}]{Jebaraj23b}
{Jebaraj}, I.~C., {Kouloumvakos}, A., {Dresing}, N., {et~al.} 2023{\natexlab{b}}, \aap, 675, A27, \dodoi{10.1051/0004-6361/202245716}

\bibitem[{{Kellogg}(1980)}]{Kellog80}
{Kellogg}, P.~J. 1980, \apj, 236, 696, \dodoi{10.1086/157789}

\bibitem[{{Kim} {et~al.}(2007){Kim}, {Cairns}, \& {Robinson}}]{Kim2007}
{Kim}, E.-H., {Cairns}, I.~H., \& {Robinson}, P.~A. 2007, \prl, 99, 015003, \dodoi{10.1103/PhysRevLett.99.015003}

\bibitem[{{Kim} {et~al.}(2008){Kim}, {Cairns}, \& {Robinson}}]{Kim2008}
---. 2008, Physics of Plasmas, 15, 102110, \dodoi{10.1063/1.2994719}

\bibitem[{{Krafft} \& {Savoini}(2022)}]{Krafft22b}
{Krafft}, C., \& {Savoini}, P. 2022, \apjl, 934, L28, \dodoi{10.3847/2041-8213/ac7f28}

\bibitem[{{Krasnoselskikh} {et~al.}(2019){Krasnoselskikh}, {Voshchepynets}, \& {Maksimovic}}]{Krasnoselskikh19}
{Krasnoselskikh}, V., {Voshchepynets}, A., \& {Maksimovic}, M. 2019, \apj, 879, 51, \dodoi{10.3847/1538-4357/ab22bf}

\bibitem[{{Krupar} {et~al.}(2020){Krupar}, {Szabo}, {Maksimovic}, {Kruparova}, {Kontar}, {Balmaceda}, {Bonnin}, {Bale}, {Pulupa}, {Malaspina}, {Bonnell}, {Harvey}, {Goetz}, {Dudok de Wit}, {MacDowall}, {Kasper}, {Case}, {Korreck}, {Larson}, {Livi}, {Stevens}, {Whittlesey}, \& {Hegedus}}]{Krupar20}
{Krupar}, V., {Szabo}, A., {Maksimovic}, M., {et~al.} 2020, \apjs, 246, 57, \dodoi{10.3847/1538-4365/ab65bd}

\bibitem[{{Leblanc} {et~al.}(1998){Leblanc}, {Dulk}, \& {Bougeret}}]{Leblanc98}
{Leblanc}, Y., {Dulk}, G.~A., \& {Bougeret}, J.-L. 1998, \solphys, 183, 165

\bibitem[{{Lecacheux}(1978)}]{Lecacheux78}
{Lecacheux}, A. 1978, \aap, 70, 701

\bibitem[{{Liu} {et~al.}(2023){Liu}, {Issautier}, {Moncuquet}, {Meyer-Vernet}, {Maksimovic}, {Huang}, {Martinovic}, {Griton}, {Chrysaphi}, {Jagarlamudi}, {Bale}, {Pulupa}, {Kasper}, \& {Stevens}}]{Liu23}
{Liu}, M., {Issautier}, K., {Moncuquet}, M., {et~al.} 2023, \aap, 674, A49, \dodoi{10.1051/0004-6361/202245450}

\bibitem[{{Maksimovic} {et~al.}(2020){Maksimovic}, {Bale}, {Chust}, {Khotyaintsev}, {Krasnoselskikh}, {Kretzschmar}, {Plettemeier}, {Rucker}, {Sou{\v{c}}ek}, {Steller}, {{\v{S}}tver{\'a}k}, {Tr{\'a}vn{\'\i}{\v{c}}ek}, {Vaivads}, {Chaintreuil}, {Dekkali}, {Alexandrova}, {Astier}, {Barbary}, {B{\'e}rard}, {Bonnin}, {Boughedada}, {Cecconi}, {Chapron}, {Chariet}, {Collin}, {de Conchy}, {Dias}, {Gu{\'e}guen}, {Lamy}, {Leray}, {Lion}, {Malac-Allain}, {Matteini}, {Nguyen}, {Pantellini}, {Parisot}, {Plasson}, {Thijs}, {Vecchio}, {Fratter}, {Bellouard}, {Lorf{\`e}vre}, {Danto}, {Julien}, {Guilhem}, {Fiachetti}, {Sanisidro}, {Laffaye}, {Gonzalez}, {Pontet}, {Qu{\'e}ruel}, {Jannet}, {Fergeau}, {Brochot}, {Cassam-Chenai}, {Dudok de Wit}, {Timofeeva}, {Vincent}, {Agrapart}, {Delory}, {Turin}, {Jeandet}, {Leroy}, {Pellion}, {Bouzid}, {Katra}, {Piberne}, {Recart}, {Santol{\'\i}k}, {Kolma{\v{s}}ov{\'a}}, {Krupa{\v{r}}}, {Krupa{\v{r}}ov{\'a}}, {P{\'\i}{\v{s}}a}, {Uhl{\'\i}{\v{r}}}, {L{\'a}n}, {Ba{\v{s}}e}, {Ahl{\`e}n},
  {Andr{\'e}}, {Bylander}, {Cripps}, {Cully}, {Eriksson}, {Jansson}, {Johansson}, {Karlsson}, {Puccio}, {B{\v{r}}{\'\i}nek}, {{\"O}ttacher}, {Panchenko}, {Berthomier}, {Goetz}, {Hellinger}, {Horbury}, {Issautier}, {Kontar}, {Krucker}, {Le Contel}, {Louarn}, {Martinovi{\'c}}, {Owen}, {Retino}, {Rodr{\'\i}guez-Pacheco}, {Sahraoui}, {Wimmer-Schweingruber}, {Zaslavsky}, \& {Zouganelis}}]{Maksimovic20}
{Maksimovic}, M., {Bale}, S.~D., {Chust}, T., {et~al.} 2020, \aap, 642, A12, \dodoi{10.1051/0004-6361/201936214}

\bibitem[{{McLean} \& {Melrose}(1985)}]{McLean85book}
{McLean}, D.~J., \& {Melrose}, D.~B. 1985, in Solar Radiophysics: Studies of Emission from the Sun at Metre Wavelengths, ed. D.~J. {McLean} \& N.~R. {Labrum} (Cambridge University Press), 237--251

\bibitem[{{Melnik} {et~al.}(2018){Melnik}, {Brazhenko}, {Frantsuzenko}, {Dorovskyy}, \& {Rucker}}]{Melnik18}
{Melnik}, V.~N., {Brazhenko}, A.~I., {Frantsuzenko}, A.~V., {Dorovskyy}, V.~V., \& {Rucker}, H.~O. 2018, \solphys, 293, 26, \dodoi{10.1007/s11207-017-1234-9}

\bibitem[{{Melrose}(1980)}]{Melrose80}
{Melrose}, D.~B. 1980, \ssr, 26, 3

\bibitem[{{Melrose}(2006)}]{Melrose2006}
---. 2006, \apj, 637, 1113, \dodoi{10.1086/498499}

\bibitem[{{M{\"u}ller} {et~al.}(2013){M{\"u}ller}, {Marsden}, {St. Cyr}, \& {Gilbert}}]{Muller13}
{M{\"u}ller}, D., {Marsden}, R.~G., {St. Cyr}, O.~C., \& {Gilbert}, H.~R. 2013, \solphys, 285, 25, \dodoi{10.1007/s11207-012-0085-7}

\bibitem[{{Page} {et~al.}(2022){Page}, {Bassett}, {Lecacheux}, {Pulupa}, {Rapetti}, \& {Bale}}]{Page22}
{Page}, B., {Bassett}, N., {Lecacheux}, A., {et~al.} 2022, \aap, 668, A127, \dodoi{10.1051/0004-6361/202244621}

\bibitem[{{Papadopoulos} \& {Freund}(1979)}]{Papadopoulos79}
{Papadopoulos}, K., \& {Freund}, H.~P. 1979, \ssr, 24, 511, \dodoi{10.1007/BF00172213}

\bibitem[{{Parker}(1960)}]{Parker60}
{Parker}, E.~N. 1960, \apj, 132, 821, \dodoi{10.1086/146985}

\bibitem[{{Pulupa} {et~al.}(2017){Pulupa}, {Bale}, {Bonnell}, {Bowen}, {Carruth}, {Goetz}, {Gordon}, {Harvey}, {Maksimovic}, {Mart{\'\i}nez-Oliveros}, {Moncuquet}, {Saint-Hilaire}, {Seitz}, \& {Sundkvist}}]{Pulupa17}
{Pulupa}, M., {Bale}, S.~D., {Bonnell}, J.~W., {et~al.} 2017, Journal of Geophysical Research (Space Physics), 122, 2836, \dodoi{10.1002/2016JA023345}

\bibitem[{{Pulupa} {et~al.}(2020){Pulupa}, {Bale}, {Badman}, {Bonnell}, {Case}, {de Wit}, {Goetz}, {Harvey}, {Hegedus}, {Kasper}, {Korreck}, {Krasnoselskikh}, {Larson}, {Lecacheux}, {Livi}, {MacDowall}, {Maksimovic}, {Malaspina}, {Mart{\'\i}nez Oliveros}, {Meyer-Vernet}, {Moncuquet}, {Stevens}, \& {Whittlesey}}]{Pulupa20}
{Pulupa}, M., {Bale}, S.~D., {Badman}, S.~T., {et~al.} 2020, \apjs, 246, 49, \dodoi{10.3847/1538-4365/ab5dc0}

\bibitem[{{Reiner} {et~al.}(2007){Reiner}, {Fainberg}, {Kaiser}, \& {Bougeret}}]{Reiner07}
{Reiner}, M.~J., {Fainberg}, J., {Kaiser}, M.~L., \& {Bougeret}, J.~L. 2007, \solphys, 241, 351, \dodoi{10.1007/s11207-007-0277-8}

\bibitem[{{Reiner} \& {MacDowall}(2019)}]{Reiner19}
{Reiner}, M.~J., \& {MacDowall}, R.~J. 2019, \solphys, 294, 91, \dodoi{10.1007/s11207-019-1476-9}

\bibitem[{{Robinson} \& {Cairns}(1998{\natexlab{a}})}]{Robinson98a}
{Robinson}, P.~A., \& {Cairns}, I.~H. 1998{\natexlab{a}}, \solphys, 181, 363, \dodoi{10.1023/A:1005018918391}

\bibitem[{{Robinson} \& {Cairns}(1998{\natexlab{b}})}]{Robinson98b}
---. 1998{\natexlab{b}}, \solphys, 181, 395, \dodoi{10.1023/A:1005033015723}

\bibitem[{{Robinson} \& {Cairns}(1998{\natexlab{c}})}]{Robinson98c}
---. 1998{\natexlab{c}}, \solphys, 181, 429, \dodoi{10.1023/A:1005023002461}

\bibitem[{{Steinberg} {et~al.}(1971){Steinberg}, {Aubier-Giraud}, {Leblanc}, \& {Boischot}}]{Steinberg71}
{Steinberg}, J.~L., {Aubier-Giraud}, M., {Leblanc}, Y., \& {Boischot}, A. 1971, \aap, 10, 362

\bibitem[{{Suzuki} \& {Dulk}(1985)}]{Suzuki85book}
{Suzuki}, S., \& {Dulk}, G.~A. 1985, in Solar Radiophysics: Studies of Emission from the Sun at Metre Wavelengths, ed. D.~J. {McLean} \& N.~R. {Labrum} (Cambridge University Press), 289--332

\bibitem[{{Tkachenko} {et~al.}(2021){Tkachenko}, {Krasnoselskikh}, \& {Voshchepynets}}]{Tkachenko21}
{Tkachenko}, A., {Krasnoselskikh}, V., \& {Voshchepynets}, A. 2021, \apj, 908, 126, \dodoi{10.3847/1538-4357/abd2bd}

\bibitem[{{Voshchepynets} \& {Krasnoselskikh}(2015)}]{Voshchep15b}
{Voshchepynets}, A., \& {Krasnoselskikh}, V. 2015, Journal of Geophysical Research (Space Physics), 120, 10,139, \dodoi{10.1002/2015JA021705}

\bibitem[{{Voshchepynets} {et~al.}(2015){Voshchepynets}, {Krasnoselskikh}, {Artemyev}, \& {Volokitin}}]{Voshchep15a}
{Voshchepynets}, A., {Krasnoselskikh}, V., {Artemyev}, A., \& {Volokitin}, A. 2015, \apj, 807, 38, \dodoi{10.1088/0004-637X/807/1/38}

\bibitem[{{Weber} {et~al.}(1977){Weber}, {Fitzenreiter}, {Novaco}, \& {Fainberg}}]{Weber77}
{Weber}, R.~R., {Fitzenreiter}, R.~J., {Novaco}, J.~C., \& {Fainberg}, J. 1977, \solphys, 54, 431, \dodoi{10.1007/BF00159934}

\bibitem[{{Zaslavsky} {et~al.}(2011){Zaslavsky}, {Meyer-Vernet}, {Hoang}, {Maksimovic}, \& {Bale}}]{Zaslavsky11}
{Zaslavsky}, A., {Meyer-Vernet}, N., {Hoang}, S., {Maksimovic}, M., \& {Bale}, S.~D. 2011, Radio Science, 46, RS2008, \dodoi{10.1029/2010RS004464}

\end{thebibliography}
\bibliographystyle{aasjournal}

\end{document}